\documentclass[]{aa}
\usepackage{amsmath}
\usepackage{latexsym}
\usepackage{longtable}
\usepackage{natbib}
\usepackage[latin1]{inputenc}
\usepackage{graphics}
\usepackage{epsfig}
\usepackage{graphicx}

\begin{document}

\title{Porosity criterion for hyperbolic voids and the cosmic microwave background}
\author{V.G.Gurzadyan\inst{1},  A.A.~Kocharyan\inst{1,2}}

\institute
{\inst{1} Yerevan Physics Institute and Yerevan State University, Yerevan,
Armenia\\
\inst{2} School of Mathematical Sciences, Monash University, Clayton, Australia
}

\date{Received (\today)}

\titlerunning{Voids and CMB}

\authorrunning{V.G.Gurzadyan and A.A.Kocharyan}

\abstract{
The degree of randomness in the cosmic microwave background (CMB) maps has been shown to be measurable and indicative of the hyperbolic effect of voids on the propagation of the photon beams. In terms of an introduced porosity parameter, we now obtain the criterion for a hyperbolicity due to domination of voids in the line-of-sight direction.
The criterion seems to be supported by the parameters of 30 Mpc scale voids revealed by the
galactic redshift surveys. The importance of the geometrical effect of the voids is also due to their possible role in  the temperature independent ellipticity of the excursion sets, as well as in the nature of dark energy.   
The role of possible larger scale inhomogeneities, from superclusters to semi-Hubble scale, in this effect remains yet unclear.   
}

\keywords{cosmology,\,\,\,cosmic background radiation}

\maketitle

\section{Introduction}

The role of matter inhomogeneities in the Universe is currently being explored with respect to
the crucial cosmological empirical phenomena: the dark energy and cosmic microwave background (CMB). The inhomogeneities might affect the observational distance scale indicators (backreaction) responsible for the introduction of dark energy, due to the distortions of the photon trajectories while propagating through the filaments and the voids (Mattsson 2007; Wiltshire 2008; Ellis 2008; Larena et al 2008). The voids are also among the
suggested reasons for non-Gaussian anomalies in the CMB; their contribution overlapped with the integrated Sachs-Wolfe effect and primordial non-Gaussianity (e.g. \cite{anomaly1,IS,HCG}). The alignment of the vectors of the CMB low multipoles and the Cold Spot, the large-scale plane-mirroring in the CMB maps, are examples of non-Gaussianities reported for WMAP data (see \cite{anomaly3,anomaly2,anomaly4}).  

The filamentary structure, i.e. the existence of voids and walls as a basic property of the large scale
distribution of galaxies, has been known for some time. The void nature of the Cold Spot is among the discussed possibilities.
 
The novelty of the current investigation is to assess whether the inhomogeneities can be treated as small scale, or whether semi-Hubble scale inhomogeneities are responsible for the anomalies.
This choice of scales is associated with a basic dichotomy, namely, whether our own position can be considered as typical (the Copernican principle) or whether we find ourselves in a particular location e.g. at the center of a void \cite{CMB_void1,CMB_void2,CMB_void3}.

CMB maps vs. large scale survey cross-correlations are currently more indicative of the integrated Sachs-Wolfe effect (see McEwen et al 2008), rather than for the voids.  It is remarkable that a 20-45\% decrease in the surface brightness and number counts of NRAO Sky Survey radio sources close to the Cold Spot has been reported \cite{Ru}, with possible association with a void, while Smith \& Huterer (2008) found no evidence for a radio flux anomaly.        

Observationally, the following parameters are attributed to typical voids: a mean scale of about 30 Mpc, a density contrast of about -1, i.e. almost absence of galaxies in the voids (e.g. \cite{Hoyle,Cec}). Larger scale filaments, supercluster-void structures, are attributed scales of the order of 100 Mpc \cite{E1,E2,E3}. In either case, obviously, the issue is in certain distribution of the void scales, rather then in a strict lattice-type structure.

Within the geometrical approach it was shown that voids can induce divergence of null geodesics and hyperbolicity \cite{GK1}, even in a globally flat or slightly positively curved Universe. The role of the voids acting as divergent lenses was noticed in Das \& Spergel (2008).  
Here the randomization in the temperature maps due to the hyperbolicity would be measurable, e.g. via Kolmogorov's stochasticity parameter 
(Gurzadyan \& Kocharyan 2008; Gurzadyan et al 2008a); the case of the Cold Spot was also considered.  Hyperbolicity can be related to the temperature independent ellipticity of the excursion sets found in CMB maps (Gurzadyan \& Torres 1997; Gurzadyan et al 2005, 2007).   

If the inhomogeneities are able to influence the dark energy effects and be traceable in the CMB properties, then it is of particular importance, first, to reveal the conditions for such effects in terms of observable descriptors, and second, to examine whether the available observational data from large scale galaxy distribution surveys support those criteria.

\section{Criterion for instability}

In the study of the properties of the photon beams in the slightly inhomogeneous Universe 
in Gurzadyan \& Kocharyan (2008a), it was shown that for the Robertson-Walker (RW) metric
\begin{equation}
ds^2 =  -(1+2\phi)\ dt^2 + (1-2\phi)\ a^2(t)\ \gamma_{mn}(x)dx^mdx^n,
\end{equation}
with perturbation $|\phi|\ll 1$, the averaged Jacobi equation i.e. the equation of deviation of close
geodesics, for the length of the deviation vector $\ell$  can be written in the form
\begin{equation}
\frac{d^2\ell}{d\lambda^2}+ r\ \ell = 0,
\end{equation}
where
\begin{equation}
\lambda(z,\Omega_\Lambda,\Omega_m)
=\int_0^z\frac{d\xi}{\sqrt{\Omega_\Lambda +[1- \Omega_\Lambda + \Omega_m \xi]\ (1+\xi)^2}}
\end{equation}
and
\begin{equation}
r = -\Omega_k +2\Omega_m\delta_0,
\end{equation}
\begin{equation}
\delta_0 \equiv \frac{\delta\rho_0}{\rho_0},
\end{equation}
$\delta_0$ is the density contrast with respect the mean density $\rho_0$, as a quantitative descriptor of inhomogeneities;  $\Omega_i, i=k, m, \Lambda$ are the usual density parameters for the curvature, matter and dark energy, respectively. 

The advantage of these formulae is that they clearly reveal the role of inhomogeneities. 
If in a purely RW Universe the scalar curvature $k=0, \pm 1$
completely defines the behavior of the propagation of photon beams, here we see that even 
when $k=0$, $\Omega_k=0$, i.e. the global curvature is zero, the density contrast affects the propagation. The latter undertakes the role of a geometry (curvature) in determining the behavior of close geodesics. This general conclusion in the particular case of underdense regions, i.e. of the voids with $\delta_{void}=(\rho_{void}-\rho_0)/\rho_0 < 0$, reveals them as hyperbolic regions deviating from the null geodesics.        
  
Let us try to find the criterion for the cumulative role of voids in an inhomogeneous Universe in terms of observable characteristic parameters of the filaments.  

First, assuming that $\Omega_k=0$, then $\Omega_\Lambda +\Omega_m = 1$, we have
\begin{equation}\label{elltau}
\frac{d^2\ell}{d\tau^2}+ \delta_0\ \ell = 0,
\end{equation}
where
\begin{equation}\label{tau}
\tau(z,\Omega_m)=\sqrt{2\Omega_m}\ \int_0^z\frac{d\xi}{\sqrt{1+\Omega_m[(1+\xi)^3-1]}}.
\end{equation}

Adopting, for simplicity, periodicity in the line-of-sight distribution of voids, i.e. $\delta_0$ periodic, $\delta_0(\tau+\tau_\kappa+\tau_\omega)=\delta_0(\tau)$ and
\begin{eqnarray}
\delta_0=
\begin{cases}
+\kappa^2\ & 0<\tau<\tau_k\ ,\\
-\omega^2\ & \tau_\kappa<\tau<\tau_\kappa+\tau_\omega\ ,
\end{cases}
\end{eqnarray}
the solution to Eq.(\ref{elltau}) 
is unstable if (see e.g. Arnold 1989) $\mu=\nu-2>0$, where
\begin{equation*}
\nu=\left|2\cos(\omega\tau_\omega)\cosh(\kappa\tau_\kappa)
+\left(\frac{\kappa}{\omega}-\frac{\omega}{\kappa}\right)
\sin(\omega\tau_\omega)\sinh(\kappa\tau_\kappa)\right|.
\end{equation*}
In our case $0<\omega\tau_\omega\ll 1$ and $0<\kappa\tau_\kappa\ll 1$. It is easy to see that if 
\begin{equation}
\bar{\delta_0} \equiv \int_0^{\tau_\kappa+\tau_\omega}\delta_0 d\tau
=\kappa^2\tau_\kappa-\omega^2\tau_\omega\ne 0,
\end{equation}
then $\mu\approx \bar{\delta_0}(\tau_\kappa + \tau_\omega)$. Thus, 
$\mu<0$, if $\bar{\delta_0}<0$, and $\mu>0$, if $\bar{\delta_0}>0$. On the other hand, if $\bar{\delta_0}=0$ then one can show that
\begin{equation}
\mu\approx -\frac{[(\kappa\tau_\kappa)^2+(\omega\tau_\omega)^2]^2}{12}<0.
\end{equation}

Then, the solutions are stable, if $\bar\delta_0\le 0$ and unstable, if $\bar\delta_0>0$, where in terms of physical descriptors 
\begin{equation}
\bar\delta_0\sim -\delta_{void}L_{void}-\delta_{wall} L_{wall};
\end{equation}
$L_{void}, L_{wall}$, $\delta_{void}, \delta_{wall}$ being the distance scales and the density contrasts of the voids and the walls, respectively.  

\section{The porosity parameter}

The available major galaxy surveys, 2 degree Field Galaxy Redshift Survey (2dFGRS), Sloan Digital Sky Survey,
Center for Astrophysics Survey, Las Campanas Redshift Survey and others, have been used to
determine the main parameters of the voids, and especially
the distribution laws of those parameters
(e.g. \cite{Hoyle,Cec} and references therein).    

One of the basic features of interest here seen in various datasets is that the voids
account for about 40\% of the total volume covered by the surveys. 
The surveys also agree on the indication, as mentioned, of the underdensity parameter of the voids 
$\delta_{void}=(\rho_{void}-\rho_0)/\rho_0 \simeq -1$. For example, the
north and south Galactic pole samples of 2dFGRS of a total of 245,591 galaxies within 1500 $\deg^2$ 
yield $-0.94\pm 0.02$ and $-0.93\pm 0.02$, respectively \cite{Hoyle}. 
The peak of the distribution of the diameters of the voids in those samples is at around 30 Mpc. 
 
If so, i.e. for the voids $\delta_{void} \simeq -1$, then the instability condition above yields
\begin{equation*}
L_{void}>\delta_{wall} L_{wall}.
\end{equation*}
This can be written as
\begin{equation}
p>
\frac{\delta_{wall}}{\delta_{wall}+1}\ , 
\end{equation}
where 
\begin{equation}\label{porosity}
p=\frac{L_{void}}{L_{wall}+L_{void}}\le 1
\end{equation}
is denoted as a parameter of {\it porosity} in the filaments.  Since the overdensity parameter of the walls then is about 2, the inequality for the porosity typically is fulfilled. 

One can easily obtain the Lyapunov exponent $\chi$
\begin{equation}
\chi = \log\left(1+\frac{\mu}{2}+\sqrt{\left(1+\frac{\mu}{2}\right)^2-1}\right)
\approx \sqrt{\mu},
\end{equation}
if $\mu\ll 1$. The mixing rate $b\approx e^{n\chi}$, where $n=\tau/(\tau_\kappa +\tau_\omega)$. 
In addition,
\begin{equation}
\mu\approx
\alpha(p-\delta_{wall}(1-p))\cdot(L_{void}+L_{wall})(\tau_\kappa + \tau_\omega),
\end{equation}
where $\alpha\approx\tau_\kappa/L_{void}\approx\tau_\omega/L_{wall}\approx(\tau_\kappa+\tau_\omega)/(L_{void}+L_{wall})$. Therefore,
\begin{equation}
\mu\approx (p-\delta_{wall}(1-p))\cdot(\tau_\kappa + \tau_\omega)^2,
\end{equation}
and
\begin{equation}
n\chi\approx
\tau\sqrt{p-\delta_{wall}(1-p)}.
\end{equation}
Thus,
\begin{equation}
b\approx e^{\tau\sqrt{p-\delta_{wall}(1-p)}}.
\end{equation}
Then the ellipticity $\epsilon$ defined as the ratio of the major and 
minor semi-axes of the excursion sets in the CMB temperature maps has the following form
\begin{equation}
\epsilon \approx e^{2n\chi} \approx e^{2\tau\sqrt{p-\delta_{wall}(1-p)}},
\end{equation}
and
\begin{equation}
p\approx\frac{1}{1+\delta_{wall}}
\left[\delta_{wall}+\left(\frac{\log\epsilon}{2\tau}\right)^2\right].
\end{equation}

The measured ellipticity for Boomerang and WMAP maps (\cite{G2,G3}), 
$\epsilon \simeq 2.2-2.5$ (depending on
the pixel counts of the excursion sets), then for $\delta_{wall} \simeq 2$ yields for the porosity parameter
\begin{equation}
p \approx \frac{2}{3} + \frac{1}{12}\cdot\left(\frac{\log\epsilon}{\tau}\right)^2 > 0.7.
\end{equation}  

\section{Conclusions}
 
The criterion for hyperbolicity in terms of the porosity parameter $p$ 
shows that the large, underdense voids ($L_{void}> L_{wall}$, $\ \delta_{void} \simeq -1$) lead to instability/hyperbolicity in the properties of null geodesics. In other words, the criterion indicates that the domination of voids
in the line-of-sight direction will support the hyperbolicity unlike the directions dominated by the walls.    
 
The comparison of the criterion with the available observational data leads to the following conclusions.
 
First, the surveys of large scale galaxy distribution indicate that the parameters of voids of 30 Mpc scale do support the condition for a hyperbolicity. 

Second, the temperature independent ellipticity of excursion sets measured for the CMB maps (Gurzadyan et al 2005, 2007), if due to the void hyperbolicity, would require a porosity parameter $p>0.7$. 

Thus, together with the degree of randomness in the CMB maps \cite{GK2}, the porosity can provide another observational link to the voids' distribution in the Universe.  

CMB maps can allow us not only to probe the parameters of the voids themselves but also the effective distance (redshift) scale of such a filamentary Universe via the 
$\tau(z,\Omega_m)$ in Eqs.(20),(21)
or e.g. to reveal the role of cosmological black holes in the formation and evolution of the voids \cite{St,Cap,Ser}.

Other inhomogeneity scales, in principle, also can contribute to this effect. This concerns the reported supercluster/void network  of around 100 Mpc (or 120 Mpc) scale \cite{E1,E2,E3}, and of larger ones, up to the Hubble scale \cite{Cec}. The properties and statistics for inhomogeneities of such larger scales in the galaxy distribution surveys, however, are still ambiguous and do not allow any conclusions on the fulfilment of the hyperbolicity condition for them. 

{\it Acknowledgments.} We are thankful to the referee for valuable comments.

\end{document}